\documentclass[12pt]{article} \textwidth=6in \oddsidemargin=0in
\begin{document}
\def\iy{\infty} \def\be{\begin{equation}} \def\ee{\end{equation}}
\def\inv{^{-1}}\def\ch{\raisebox{.4ex}{$\chi$}}
\def\ov{\over} \def\s{\sigma} \def\t{\tau} \def\a{\alpha} \def\x{\xi}
\newcommand{\twotwo}[4]{\left(\begin{array}{cc}#1&#2\\&\\#3&#4\end{array}\right)} \def\ph{\varphi} 
\def\R{\bf R} \def\d{\det\,} \def\hf{{{1\ov2}}} \def\dl{\delta} \def\Om{\Omega} \def\chpm{\ch_{(-\a,\,\a)}} \def\chp{\ch_{(0,\,\a)}}
\def\chm{\ch_{(-\a,\,0)}} \def\cp{\ch^+} \def\cm{\ch^-}
\def\om{\omega} \def\d{\det\,} \def\noi{\noindent} \def\s{\sigma}
\def\t{\tau} \def\W{W_\a} \def\x{\xi} \def\tl{\tilde} \def\tn{\otimes}
\def\G{\Gamma} \def\ps{\psi} \def\tt{\tl\t} 
\def\r{\rho} \def\ep{\varepsilon} \def\bR{{\bf R}}
\def\la{\lambda} \def\({\left(} \def\){\right)} \def\eq{\equiv}
\renewcommand{\Im}{{\rm Im}\,} \def\l{\lambda} \renewcommand{\Re}{{\rm Re}\,} \def\wt{\widetilde} \def\Oda{O(e^{-\dl\a})} \def\ga{\gamma}
\def\th{\theta}
\begin{center}{ \large\bf  Asymptotics of a Class of Operator Determinants\\ with Application to the Cylindrical Toda Equations}\end{center}

\begin{center}{\bf Harold Widom}\end{center}

\begin{center}{Department of Mathematics\\
University of California\\
Santa Cruz, CA 95064}
\end{center}
  
\renewcommand{\theequation}{1.\arabic{equation}}

\begin{center}{\bf  I. Introduction}\end{center}

The cylindrical Toda equations are
\[q_k''(t)+t^{-1}q_k'(t)=4\,(e^{q_k(t)-q_{k-1}(t)}-e^{q_{k+1}(t)-q_k(t)})\ \ \ \ (k\in{\bf Z}).\]

A class of soultions of these equations was found in \cite{W1}: 
Define the integral kernels $K_k(u,v)$ acting on $L^2(\bR^+)$ by
\be\int{e^{-t[(1-\om)u+(1-\om\inv)u\inv]}\over-\om u+v}\,\om^k\,d\rho(\om),\label{Kk}\ee
where $\r$ is any finite complex measure supported on a compact subset of 
\[\{\om\in{\bf C}:\,{\rm Re}\;\om<1,\ {\rm Re}\;\om\inv<1\}.\]
Then 
\[q_k(t)=\log\,\det\,(I+K_k)-\log\,\det\,(I+K_{k-1})\]
is a solution to the system of equations. If $\r$ is supported in the set of $n$th roots of unity then the solution is $n$-periodic in that $q_{k+n}=q_k$ for all $k$.
The main interest is in the limit $t\to0+$, while the $t\to\iy$ asymptotics are straightforward. 

The asymptotics for the $q_k$ are known once those for the determinants are, and since $\r$ is arbitrary there is no loss of generality if one considers $\d(I+K_0)$ only.
The asymptotics for this were determined in \cite{TW} in what we call here the regular case. This is where the function
\be h(s)=1+{\pi\ov\sin\pi s}\,\int (-\om)^{s-1}\,d\r(\om)\label{h}\ee
satisfies
\be h(s)\ne0\ {\rm for\ Re}\,s=\hf,\ \ \ \Delta\,{\rm arg}\;h(s)\Big|_{\hf-i\iy}^{\hf+i\iy}=0.\label{hreg}\ee
The asymptotics were of the form
\be\d(I+K_0)\sim b\,t^a\label{regasym}\ee
where $a$ and $b$ were given by integral formulas. In the periodic case they were given in terms of the zeros of $h(s)$ by formulas (\ref{a}) and (\ref{b}) below. The derivation was rather intricate, involving the computation of resolvent operators.\footnote{The result of \cite{TW} is in one sense less general and in another sense more general than stated here, but this is not significant.}
 
In \cite{W2} we found asymptotic formulas for a class of operator determinants depending on a parameter generalizing these, but having some of the same qualitative properties. The asymptotics were of the expected form
$b\,t^\a$,
the constant $a$ was given by an integral formula, but $b$ was itself given in terms of an operator determinant. This is probably all that one can say in general but from \cite{TW} we know that for the Toda kernels, as we shall call (\ref{Kk}), there is a more explicit representation. 

The method of \cite{W2} deals with determinants directly rather than resolvent operators. Although it treated the regular case only it raised the possiblility of carrying the method further to handle some singular cases. This is what we do in the present work. For these singular cases the symbol will have a double zero at a single point on the line, although the method could likely be extended to other cases as well. The asymptotics take the form
$b\,t^\a\,\log t\inv$.

We state here the results for the $n$-periodic Toda kernels $K_0$, since these are the most interesting cases from the point of view of integrable systems. From periodicity it follows that in the regular case $h(s)$ has $n$ zeros the strip $\hf<{\rm Re}\,s<n+\hf$, say at $\a_1,\ldots,\a_n$. In the result of \cite{TW} the constants in (\ref{regasym}) are given by
\be a={1\ov n}\,\sum \a_k^2-{(n+1)(2n+1)\ov6},\label{a}\ee
\be b=n^{-a}\,{\prod_{j,k}}{G\({j-k\ov n}+1\)\ov G\({\a_j-\a_k\ov n}+1\)},\label{b}\ee
where $G$ is the Barnes $G$-function. (This is an entire function satisfying the functional equation $G(z+1)=\G(z)\,G(z)$.) The quotient of the two sides in the asymptotics is $1+O(t^\dl)$ for some $\dl>0$.

In the singular case we consider here $h(s)$ has a double zero at $s=\hf$. (The exact condition will be stated in \S III.1.) In this case we set $\a_1=\hf$ and $\a_n=n+\hf$. The other zeros, $\a_2,\ldots,\a_{n-1}$, lie inside the strip. The asymptotics become now
\[\d(I+K_0)\sim b\,t^a\,\log t\inv,\]
with $a$ as before but
\[b= 2\,n^{1-a}\,{\prod_{j,k}}'{G\({j-k\ov n}+1\)\ov G\({\a_j-\a_k\ov n}+1\)},\]
where $\prod'$ indicates that the factor in the denominator corresponding to $j=1,\ k=n$ is omitted. The quotient of the two sides in the asymptotics is of the form
\[1+p_1/(\log t\inv)+p_2/(\log t\inv)^2+\cdots\]
times $1+O(t^\dl)$, where the series is asymptotic and convergent for sufficiently small $t$, and
\be p_1=-{2\ov n}\,\ga+\log n
+{1\ov 2n}\sum_{k\ne1,n}\left[{\G'\ov\G}\!\({\a_k-\hf\ov n}\)+{\G'\ov\G}\!\({\hf-\a_k\ov n}+1\)\right].\label{p1}\ee

So as not to require the reader to be familiar with \cite{W2} or \cite{TW} we begin at the beginning, with the regular case for the more general family of kernels. Then we go further than in \cite{W2} in that we rederive the integral representation for the constants for the Toda kernels, and also the formulas (\ref{a}) and (\ref{b}) in the periodic subcase. The derivations will be different from those in \cite{TW}. 

The section following will treat the singular cases. What will allow us to use essentially the same method are asymptotic formulas for truncated Wiener-Hopf operators with symbol having a Fisher-Hartwig singularity (the simplest), which were obtained for this purpose \cite{W3}.

Finally we apply these results in two special cases, the cylindrical sinh-Gordon equation and the Bullough-Dodd equation, where asymptotics have already been derived \cite{MTW,K}, but for which rigourous proofs have not (as far as we know) appeared in the litearture.

\setcounter{equation}{0}\renewcommand{\theequation}{2.\arabic{equation}}
 
\begin{center}{\bf II. The regular case}\end{center}

If one makes the substitutions $u\to e^x,\ v\to e^y$ in (\ref{Kk}) with $k=0$, and sets $t=e^{-\a}$, then the operator $K_0$ on $L^2(\R^+)$ becomes the operator on $L^2(\R)$ with kernel
\be K_\a(x,y)=\int{e^{-[(1-\om) e^{x-\a}+(1-\om\inv) e^{-x-\a}]}\ov
-\om\, e^{(x-y)/2}+e^{(y-x)/2}}d\r(\om).\label{todakernel}\ee
As $\a\to\iy$ this kernel has the pointwise limit
\[K(x,y)= \int {1\ov-\om \,e^{(x-y)/2}+e^{(y-x)/2}}d\r(\om).\]
The convergence is not uniform throughout $\bR$. But for $x\in\bR^+$ the second summand in the exponent converges uniformly to zero, with the result that the kernel converges uniformly to a translate of the kernel
\[K_+(x,y)= \int {e^{-(1-\om) \,e^x}\ov-\om\, e^{(x-y)/2}+e^{(y-x)/2}}d\r(\om).\]
Similarly, for $x\in\bR^-$ the kernel converges uniformly to a translate of the kernel
\[K_-(x,y)= \int {e^{-(1-\om\inv)\, e^{-x}}\ov-\om \,e^{(x-y)/2}+e^{(y-x)/2}}d\r(\om).\]

We isolate some of the properties of the family $K_\a$ with its three associated operators $K,\ K_\pm$, and derive asymptotics as $\a\to\iy$ in a  general setting. To obtain the asymptotics of $\d(I+K_0)$ in the notation of (\ref{Kk}) from the asymptotics for $\d(I+K_\a)$ in the notation of (\ref{todakernel}) one need only set $\a=\log t\inv$.

\begin{center}{\bf 1. Statement of the result}\end{center}

The setting is a family of trace class operators $K_\a$ on $L^2(\R)$ with three associated operators $K,\ K_\pm$. The operator $K$ has kernel $k(x-y)$ where $k\in L^1(\R)$. 

The first connection between the operators $K_\a$ and $K$ is that
\be\ch^-\,(K_\a-K)\,\ch^+=o_1(1),\ \ \ch^+\,(K_\a-K)\,\ch^-=o_1(1).
\label{cond0}\ee
Here $\ch^\pm$ denotes multiplication by $\ch_{\R^\pm}$ and $o_1(1)$ denotes any family of operators whose trace norms are $o(1)$. (In particular, $\ch^-\,K\,\ch^+$ and $\ch^-\,K\,\ch^+$ are trace class.)

The translation operator $T_a$ is defined by $T_a f(x)=f(x-a)$. The second main assumption is that there are operators $K_\pm$ such that
\be\ch^+(\,K_\a-T_\a\,K_+\,T_{-\a})\,\ch^+=o_1(1),\ \ \ 
\ch^-\,(K_\a-T_{-\a}\,K_-\,T_{\a})\,\ch^-=o_1(1),\label{cond1}\ee 
and such that
\be L_-:=K_--\ch^+\,K\,\ch^+\ \ {\rm and}\ \ L_+:=K_+-\ch^-\,K\,\ch^-
\ \ {\rm are\ trace\ class}.\label{cond2}\ee
 
What characterizes the regular case is that if $\hat k$ is the Fourier transform of $k$,
\[\hat k(\x)=\int_{-\iy}^\iy e^{ix\x}\,k(x)\,dx,\]
and if $\s(\x)=1+\hat k(\x)$, then\footnote{For the Toda operators, if one makes the variable change $s=\hf+i\x$ then $\s(\x)$ becomes the function $h(s)$ given by (\ref{h}), and (\ref{index}) is the same as  (\ref{hreg}).}
\be \s(\x)\ne0,\ \ \ {\rm arg}\;\s(\x)|_{-\iy}^\iy=0.\label{index}\ee
This assures that the Wiener-Hopf operators $W^\pm(\s)=I+\ch^\pm\,K\,\ch^\pm$ are invertible.\footnote{See, for example, \cite[\S I.8]{GF}. Wiener-Hopf operators traditionally act on $L^2({\bf R})$, but $W^\pm(\s)$ will act on $L^2({\bf R})$.} We also assume that $|x|^{1/2}\,k(x)\in L^2(\R)$ which, together with (\ref{index}), allowed us to use the Kac-Achieser theorem.\footnote{This says that if $W_\a(\s)=I+\chpm\,K\,\chpm$ acting on $L^2(-\a,\a)$ then 
$\d W_\a(\s)\sim G(\s)^{2\a}\,E(\s)$, where $G(\s)=\exp\,(s(0)),\ E(\s)=\exp\,\{\int_0^\iy x\,s(x)\,s(-x)\,dx\}$, and $s(x)$ is the inverse Fourier transform of $\log \s(\x)$.}

The result of \cite{W2} was that if (\ref{cond0})--(\ref{cond2}) are satisfied and (\ref{index}) holds then
\be\det(I+K_\a)\sim G(\s)^{2\a}\,E(\s)\;
\d(I+W^+(\s)\inv\,L_-)\;\d(I+(I+W^-(\s)\inv\,L_+),\label{result}\ee
where $G(\s)$ and $E(\s)$ are the constants in the Kac-Achieser theorem. 

For the Toda kernels the conditions (\ref{cond0})--(\ref{cond2}) are satisfied and so (\ref{result}) holds.

\begin{center}{\bf 2. Proof of the result}\end{center}

We think of $L^2(\R)$ as $L^2(\R^-)\oplus L^2(\R^+)$ and the corresponding matrix representation of $K_\a$. Condition (\ref{cond1}) tells us that with error $o_1(1)$ this equals 
\[\twotwo{\cm\,T_{-\a}\,K_-\,T_{\a}\,\cm}{\cm\,K\,\cp}{\cp\,K\,\cm}{\cp\,T_\a\,K_+\,T_{-\a}\,\cp}.\]

Since the upper-right corner is trace class, if we multiply it on either side by $\ch_{(-\iy,\,-\a)}$ or $\ch_{(\a,\,\iy)}$ the result is $o_1(1)$. Similarly for the lower-left corner. Hence with this error the above equals
\be\twotwo{\chm\,K\,\chm}{\chm\,K\,\chp}{\chp\,K\,\chm}
{\chp\,K\,\chp}+
\twotwo{\cm\,T_{-\a}\,L_-\,T_{\a}\,\cm}{0}{0}
{\cp\,T_{\a}\,L_+\,T_{-\a}\,\cp}.\label{Kaappr}\ee

The operator
\[I+\twotwo{\chm\,K\,\chm}{\chm\,K\,\chp}{\chp\,K\,\chm}
{\chp\,K\,\chp}\]
is just $W_\a(\s)$ in its matrix representation and so its determinant is asymptotically equal to $G(\s)^{2\a}\,E(\s)$, by the Kac-Achieser theorem.\footnote{Strictly speaking $W_\a(\s)$ acts on $L^2(-\a,\a)$ while the above operator acts on $L^2(\R)$. But the determinants are the same.}

The next step is to factor out this operator from $I+K_\a$
and determine the asymptotics of the determinant of the result. Thus 
$\d(I+K_\a)$ equals $\d\W(\s)$ times the determinant of $I$ plus
\be\twotwo{\cm\,W_\a\inv\,\cm}{\cm\,W_\a\inv\,\cp}{\cp\,W_\a\inv\,\cm}{\cp\,W_\a\inv\,\cp}\,\twotwo{\cm\,T_{-\a}\,L_-\,T_{\a}\,\cm}{0}{0}
{\cp\,T_{\a}\,L_+\,T_{-\a}\,\cp}+o_1(1).\label{Kaappr2}\ee
The reason the error remains $o_1(1)$ after multiplying by $W_\a(\s)\inv$ is that these operators have uniformly bounded norms.
(See \cite[\S III.1]{GF}.)

Without change of notation or determinant we may think of our operators as acting on $L^2(\R)\oplus L^2(\R)$, and then the determinant of $I$ plus the above is the same if the product is replaced by

\[\twotwo{T_{\a}}{0}{0}{T_{-\a}}
\twotwo{\cm\,W_\a\inv\,\cm}{\cm\,W_\a\inv\,\cp}
{\cp\,W_\a\inv\,\cm}{\cp\,W_\a\inv\,\cp}\]
\be\times \twotwo{\cm\,T_{-\a}\,L_-\,T_{\a}\,\cm}{0}{0}
{\cp\,T_{\a}\,L_+\,T_{-\a}\,\cp}\twotwo{T_{-\a}}{0}{0}{T_{\a}}.\label{prod}\ee

The two operators involving $L_-$, which will end up in the first column, are 
\be T_\a\,\cm\,W_\a\inv\,\cm\,T_{-\a}\,L_-\,T_{\a}\,\cm\,T_{-\a}\ \ {\rm and}
\ \ T_{-\a}\,\cp\,W_\a\inv\,\cm\,T_{-\a}\,L_-\,T_{\a}\,\cm\,T_{-\a}.\label{twoops}\ee
If we use the general fact
\[T_\a\,\ch_J=\ch_{J+\a}\,T_\a\]
we see that these are 
\[\ch_{(-\iy,\a)}\,(I+\ch_{(0,2\a)}\,K\,\ch_{(0,2\a)})\inv\,\ch_{(-\iy,\a)}\,L_-\,\ch_{(-\iy,\a)}\]
\be=\cm\,L_-\,\ch_{(-\iy,\a)}+\ch_{(0,\a)}(I+\ch_{(0,2\a)}\,K\,\ch_{(0,2\a)})\inv\,\ch_{(0,\a)}\,L_-\,\ch_{(-\iy,\a)}\label{first}\ee
and
\[T_{-\a}\,\ch_{(0,\a)}\,(I+\ch_{(-\a,\a)}\,K\,
\ch_{(-\a,\a)})\inv\,\ch_{(-\a,0)}\,T_{-\a}\, L_-\,\ch_{(-\iy,\a)}\]
\be=\ch_{(-\a,0)}\,T_{-2\a}\,(I+\ch_{(0,2\a)}\,K\,\ch_{(0,2\a)})\inv\,
\ch_{(0,\a)}\,L_-\,\ch_{(-\iy,\a)}.\label{second}\ee
The other entries of the matrix kernels are analogous to these.

Since $L_-$ is trace class, $L_-\,\ch_{(-\iy,\a)}$ converges to $L_-$ in trace norm as $\a\to\iy$. Furthermore $(I+\ch_{(0,2\a)}\,K\,\ch_{(0,2\a)})\inv$ converges strongly to $(I+\cp\,K\,\cp)\inv$. (For this fact also, see \cite[\S III.1]{GF}.) We deduce that (\ref{first}) converges in trace norm to
\[\cm\,L_-+\cp\,(I+\cp\,K\,\cp)\inv\,\cp\,L_-=(I+\cp\,K\,\cp)\inv\,L_-=W^+(\s)\inv\,L_-.\]

As for (\ref{second}), the factor $(I+\ch_{(0,2\a)}\,K\,\ch_{(0,2\a)})\inv\,
\ch_{(0,\a)}\,L_-\,\ch_{(-\iy,\a)}$ converges in trace norm, as above, but $\ch_{(-\a,0)}\,T_{-2\a}$ converges strongly to zero. Hence the product converges in trace norm to zero.

Similarly the upper-right entry of (\ref{prod}) converges in trace norm to zero and the lower-right to $W^-(\s)\inv\,L_+$. Thus (\ref{prod}) itself converges in trace norm to
\[\twotwo{W^+(\s)\inv\,L_-}{0}{0}{W^-(\s)\inv\,L_+},\]
and so the determinant of $I$ plus (\ref{prod}) converges to the product of determinants on the right side of (\ref{result}), which is therefore established.
 
\begin{center}{\bf 3. The Toda operators}\end{center}

It was mentioned earlier that for the Toda operators , if we set $s=\hf+i\x$ then $\s(\x)$ becomes the function $h(s)$ given by (\ref{h}), and condition (\ref{index}) is the same as (\ref{hreg}). In the following discussion we use either $\s(\x)$ or $h(s)$, whichever is more convenient.

\begin{center}{\bf a. Evaluation of the determinants in (\ref{result}) }\end{center}

We first make another assumption which we later remove. We introduce a parameter $\l$ and repace $K$ and $K^\pm$ by $\l\,K$ and $\l\,K_\pm$. Then we have the corresponding symbol $\s_\l(\x)=1+\l\,\hat{k}(\x)$. We assume temporarily that
there is a curve in $\bf C$ joining $\l=0$ to $\l=1$ so that each $\s_\l$ satisfies (\ref{index}).

Let us write
\[D^\pm(\s)=\det(I+W^\pm(\s)\inv\,L_\mp),\]
and correspondingly
\[D^\pm(\s_\l)=\det(I+\l\,W^\pm(\s_\l)\inv\,L_\mp).\]
We shall compute the logarithmic derivatives
\[{d\ov d\l}\log D^\pm(\s_\l).\]
Integrating from $\l=0$ to $\l=1$ will then give us $\log D^\pm(\s)$. 

We first compute the logarithmic derivative of $D^+(\s_\l)$. The operator in question is
\[I+\l\,W^+(\s_\l)\inv\,L_-=(I+\l\,\cp K\cp)\inv(I+\l\,K_-).\]
The logarithmic derivative of the determinant of this is the trace of
\[(I+\l\,K_-)\inv\,K_--(I+\l\,\cp\,K\,\cp)\inv\,\cp\,K\,\cp\]
\be=\l\inv\,[(I+\l\,\cp\,K\,\cp)\inv-(I+\l\,K_-)\inv].\label{opdiff}\ee.

In computing the trace we first make the assumption that $\Om$, the support of $\r$, lies strictly in the left half-plane. Then the trace is unchanged if we multiply the kernel of $K_-$ by $e^{e^{-x}-e^{-y}}$, so that it becomes
\[e^{{x+y\ov2}}\int {e^{\,\om\inv\,e^{-x}-e^{-y}}\ov-\om e^x+e^y}d\r(\om)=e^{-{x+y\ov2}}\int {e^{\,\om\inv\,e^{-x}-e^{-y}}\ov\om\inv e^{-x}-e^{-y}}\om\inv d\r(\om).\]
For convenience we shall use the same notation $K_+(x,y)$ for this kernel. We compute
\be K_-(x,y)=e^{-{x+y\ov2}}\int (-w)\inv d\r(\om)\int_0^{\iy} e^{e^{u}\,
[\om\inv e^{-x}-e^{-y}]}\,e^{u}\,du\label{K+int}\ee
\[=\int_0^{\iy}\left[\int e^{{u-x\ov2}}\,e^{\om\inv\,e^{u-x}}\,(-\om)\inv\,d\r(\om)\right]\cdot\left[e^{{u-y\ov2}}\,e^{-e^{u-y}}\right]\,du.\]
If we write $A(x,u)$ for the expression in the first bracket and $B(u,y)$ for the expression in the second, and $A$ and $B$ the correspponding operators on $L^2(\R)$, then 
$K_-=A\,\cp\,B$.
To compute the kernel of $BA=AB$ (the operators commute since they are both convolutions) we replace the lower limit $0$ in (\ref{K+int}) by $-\iy$  and find that it is precisely equal to $K(x,y)$. Thus
\[\l\inv\,[(I+\l\,\cp\,K\,\cp)\inv-(I+\l\,K_-)\inv]=\l\inv\,[(I+\l\,\cp\,BA\,\cp)\inv-(I+\l\,A\,\cp\,B)\inv].\]
If we set
\[U=\l\,A\,\cp,\ \ \ V=\cp\,B,\]
then this equals
\[\l\inv\,[(I+VU)\inv-(I+UV)\inv]=\l\inv\,[U\,(I+VU)\inv V-VU\,(I+VU)\inv]\]
\[=\l\inv\,[U\,(I+VU)\inv,\,V]=[A\,\cp\,(I+\l\,\cp\,BA\,\cp)\inv,\,\cp\,B],\]
where the last two sets of brackets indicate commuator.

Now notice the following. If we replace the last $\cp\,B$ by $\cp\,B\,\cp$ the error is trace class and so the trace of the commutator will not change. Similarly for the $A\,\cp$, and so all our operators may be thought of as acting on $L^2(\R^+)$ and we can use Wiener-Hopf formulas. 

The commutator in question is
\[[W(a)\,W(1+\l\,ab)\inv,\,W(b)],\]
where
\[a(\x)=\int (-\om)\inv\,d\r(\om)\int e^{-{x\ov2}}\,e^{\,-\om\inv\,e^{-x}}\,e^{i\x x}\,dx\]
\[=\int (-\om)\inv\,d\r(\om)\int z^{-{1\ov2}}\,e^{\,\om\inv\,z}\,z^{-i\x}\,dz
=\int (-\om)^{-i\x-{1\ov2}}\,d\r(\om)\;\G({\scriptstyle\hf}-i\x),\]
\[b(\x)=\int e^{{x\ov2}}\,e^{-e^{x}}\,e^{i\x x}\,d\x=\int z^{-{1\ov2}}\,e^{-z}\,z^{i\x}\,dz=\G({\scriptstyle\hf}+i\x).\]
Since $W(a)\,W(1+\l\,ab)\inv-W(a\,(1+\l\,ab)\inv)$ is trace class, the trace of the commuator is unchanged if the commuator is replaced by 
\[\left[W\left({a\ov1+\l\,ab}\right),\,W(b)\right].\]
We use the general fact that
\[{\rm tr}\,[W(\s_1),\,W(\s_2)]={1\ov2\pi i}\int_{-\iy}^\iy \s_1(\x)\,\s_2'(\x)\,d\x\]
and find that the trace of our commutator is
\[{1\ov2\pi i}\int_{-\iy}^\iy \left({a(\x)\ov1+\l\,a(\x)\,b(\x)}\right)\,b'(\x)\,d\x.\]
This the logarithmic derivative of $D^+(\s_\l)$.

Recall that this formula was derived under the assumption that $\Om$ lies in the left half-plane.  
To remove this condition let us assume at first that $\l$ is small, and for any $\eta$ define
the measure $\rho_{\eta}$ by $\rho_{\eta}(E)=\rho(E-\eta)$. This has support $\Om+\eta$.
For all $\eta$ in a neighborhood in ${\bf C}$ of $[-1,\,0]$ the set $\Om+\eta$ is contained in 
the region $\{\om\in{\bf C}:\,{\rm Re}\;\om<1,\ {\rm Re}\;\om\inv<1\}$  and we have corresponding operators. If $\la$ is small enough the symmbol $\s_\l$ (which now also depends on $\eta$) will satisfy (\ref{index}) for all $\eta$.
For $\eta$ near $-1$ the set $\Om+\eta$ will also lie in the left half-plane and so the formula for the logarithmic derivative holds. Since the logarithmic derivative is an analytic function of $\eta$ the formula holds for all $\eta$, in particular for $\eta=0$. Thus we have removed the condition on $\Om$ if $\l$ is small enough. Since the logarithmic derivative is analytic in $\l$ in a neighborhood of the curve, the formula holds for all $\l$ on the curve.

If we integrate the logarithmic derivative from 0 to 1 we obtain the formula
\[\log\,D^+(\s)={1\ov2\pi i}\int_{-\iy}^\iy {b'(\x)\ov b(\x)}\,\log(1+a(\x)\,b(\x))\,d\x={1\ov2\pi }\int_{-\iy}^\iy {\G'(\hf+i\x)\ov\G(\hf+i\x)}\,\log\s(\x)\,d\x.\]
One can check that the analogous integral representing $\log\,D^-(\s)$ is obtained by replacing $d\r(\om)$ by $-\om\,d\r(\om\inv)$, or equivalently by replacing $\s(\x)$ by $\s(-\x)$. Thus
\be D^-(\s)\,D^+(\s)=\exp\left\{{1\ov2\pi }\int_{-\iy}^\iy {\G'(\hf+i\x)\ov\G(\hf+i\x)}\,\log\,(\s(\x)\,\s(-\x))\,d\x\right\}.\label{detprodrep}\ee

We now remove the assumption made at the beginning of the section, that there is a curve in $\bf C$ joining $\l=0$ to $\l=1$ so that each $\s_\l$ satisfies (\ref{index}). All we use, in fact, is that (\ref{detprodrep}) holds as long as $|\s(\x)-1|<1$, that $\s=1+ab$ where $a,\,b\in C_0\cap L^2$ and that (\ref{index}) holds.  We shall show that (\ref{detprodrep}) holds whenever (\ref{index}) is satisfied and $b\ne0$, as is the case in hand when  $b(\x)=\G(\hf+i\x)$. 

Because of (\ref{index}) we can write $\s=e^\ph$ where $\ph(\pm\iy)=0$. 
Let us define
\[\s_\mu=e^{\mu\ph},\ \ \ a_\mu={e^{\mu\ph}-1\ov b},\]
so $\s_\mu=1+a_\mu\,b$. We have $a_\mu(\x)\sim\mu\,\ph(\x)/b(\x)\sim \mu\,a(\x)$ as $\x\to\iy$, so $a_\mu\in C_0\cap L^2$. Now if $\mu$ is small then $|\s_\mu-1|<1$ and so (\ref{detprodrep}) holds. But both sides of (\ref{detprodrep}) are entire functions of $\mu$ and so the identity holds for all $\mu$, including $\mu=1$. Thus (\ref{detprodrep}) holds in all cases.

\begin{center}{\bf b. Evaluation of $G(\s),\ E(\s)$, and  $D^\pm(\s)$  in the periodic case}\end{center} 

We begin with $G(\s)$. We have
\[\s(\x)=1+{\pi\ov\sin\pi (\hf+i\x)}\,\int (-\om)^{-i\x-{1\ov2}}\,d\r(\om)\]
and
\[\log G(\s)={1\ov2\pi}\int_{-\iy}^{\iy}\log \s(\x)\,d\x.\]
In terms of $h$ this becomes
\[\log G(\s)={1\ov2\pi i}\int_{\hf-i\iy}^{\hf+i\iy}\log h(s)\,ds
=-{1\ov2\pi i}\int_{\hf-i\iy}^{\hf+i\iy}{h'(s)\ov h(s)}s\,\,ds.\]

In the $n$-periodic case, where $\r$ is supported in the set of $n$th roots of unity, $h(s)$ has period $n$. Let $C_n$ be the contour running from $\hf-i\iy$ to $\hf+i\iy$ and then from $-n+\hf+i\iy$ to $-n+\hf-i\iy$. By periodicity the index of $h$ over the contour $C_n$ is zero. Since inside the contour $h$ has poles at $j=1,\ldots,n$ it must also have $n$ zeros, say at $\a_1,\ldots,\a_n$. (This was stated in the introduction.) 

Consider
\[{1\ov2\pi i}\int_{C_n}{h'(s)\ov h(s)}(s-n/2)^2\,\,ds.\]
This equals on the one hand $-2n\,\log G(\s)$ and on the other hand
\[\sum [(\a_k-n/2)^2-(k-n/2)^2]=\sum \a_k^2-n\sum(\a_k-k)-{n(n+1)(2n+1)\ov6}.\]
Considering
\[{1\ov2\pi i}\int{h'(s)\ov h(s)}s\,\,ds\]
over the same contour (using the fact that $\int_{\hf-i\iy}^{\hf+i\iy} h'(s)/h(s)\ ds=0$) gives $\sum\a_k=\sum k$. Thus 
\be 2\log G(\s)=-a,\label{G}\ee
where $a$ is given by (\ref{a}).

We next evaluate $E(\s)$ in the periodic case. The general formula is
\[E(\s)={1\ov 2\pi i}\int_{-\iy}^{\iy}\log \s(\x)\,d\x\,
{1\ov 2\pi i}\int_{-\iy-0i}^{\iy-0i}{\s'(\eta)\ov\s(\eta)}\,{d\eta\ov\eta-\x}\,d\eta.\]
We make the substitutions $s=\hf-i\x,\ t=\hf-i\eta$ and find that in terms of $h$
\[E(\s)={1\ov 2\pi i}\int_{\hf-\iy}^{\hf+\iy}\log h(s)\,ds\,
{1\ov 2\pi i}
\int_{-i\iy+\hf-}^{i\iy+\hf-}{h'(t)\ov h(t)}\,{dt\ov t-s}\,dt.\]

The inner integral is what is obtained by integrating
\[{1\ov n}{h'(t)\ov h(t)}\;{\G'\ov\G}\!\({t-s\ov n}\)\]
over $C_n$ shifted infinitesimally to the left. The second factor is analytic inside the contour so the value of the integral (with its external factor) is
\[n\,\sum_j\left[{\G'\ov\G}\!\({\a_j-s\ov n}\)-{\G'\ov\G}\!\({j-s\ov n}\)\right].\]

Therefore the next step is to evaluate
\be{1\ov 2\pi i}\int_{\hf-\iy}^{\hf+\iy}\log h(s)\ {\G'\ov\G}\!\({t-s\ov n}\)\,ds,\label{Gint}\ee
where $\hf<|\Re t|<n+\hf$. (Either $t=j$ or $t=\a_j$.) We do this by replacing the integrand by
\[\log h(s)\ {G'\ov G}\!\({t-s\ov n}\),\]
where this $G$ is the Barnes $G$-function,
and integrating over the contour $C_n-n$. By the periodicity of $h$ and the functional equation for $G$ the result is the negative of (\ref{Gint}) and integration by parts shows (\ref{Gint}) equals the integral of
\[-{1\ov n}\,{h'(s)\ov h(s)}\;\log G\({t-s\ov n}\).\]
The poles of $h$ inside the contour are now at $-n+k$ and the zeros are at $-n+\a_k$ with $k=1,\ldots,n$ as before. The second factor is analytic inside the contour. We deduce that (\ref{Gint}) equals
\[{1\ov n}\sum_k\left[\log G\({t-k\ov n}+1\)-\log G\({t-\a_k\ov n}+1\)\right].\]

Putting these things together gives
\be E(\s)=\prod_{j,k}{G\({\a_j-k\ov n}+1\)\ov G\({\a_j-\a_k\ov n}+1\)}
{G\({j-\a_k\ov n}+1\)\ov G\({j-k\ov n}+1\)}.\label{E}\ee

Finally we evaluate the product of determinants in the periodic case. In terms of $h$ identity (\ref{detprodrep}) becomes
\[\log (D^-(\s)\,D^+(\s))={1\ov2\pi i}\int_{\hf-i\iy}^{\hf+i\iy} {\G'(s)\ov\G(s)}\,\log\,(h(1-s)\,h(s))\,ds.\]

Let us first evaluate
\be {1\ov 2\pi i}\,\int_{\hf-i\iy}^{\hf+i\iy}\,{\G'(s)\ov\G(s)}\,\log h(s)\,ds.\label{Gintegral}\ee
We have
\[{\G'(s)\ov\G(s)}=\log n+{d\ov ds}\log\prod{\G\({s-k\ov n}+1\)}.\]
Integration by parts shows that the contribution of the last term is
\be -{1\ov2\pi i}\,\int_{\hf-i\iy}^{\hf+i\iy}\,\log\,\prod \G\({s-k\ov n}+1\)\,{h'(s)\ov h(s)}\,ds.\label{Gintegral2}\ee
We evaluate this by integrating
\[-{1\ov2\pi i}\log\,\prod G\({s-k\ov n}+1\)\,{h'(s)\ov h(s)}\]
over $C_n$. By periodicity of $h$ and the functional equation for $G$ the integral equals (\ref{Gintegral2}). Inside the contour the first factor is analytic and we see that the integral equals
\[\sum_{j,k}\log\,{G\({j-k\ov n}+1\)\ov G\({\a_j-k\ov n}+1\)}.\]

To evaluate the corresponding integral with $h(s)$ replaced by $h(1-s)$ we observe that inside $C_n$ this function has zeros at $n+1-\a_j$ and poles at $n+1-j$. In the resulting double sum we make the substitution $k\to n+1-k$ and find the resut to be
\[\sum_{j,k}\log\,{G\({k-j\ov n}+1\)\ov G\({k-\a_j\ov n}+1\)}.\]

Finally there is the $\log n$ appearing twice, which contributes
\[2\,\log n\,{1\ov2\pi i}\int_{\hf-i\iy}^{\hf+i\iy}\,\log h(s)\,ds=2\,\log n\,\log G(\s)=-a\,\log n.\]
Hence we obtain the final result
\be D^-(\s)\,D^+(\s)=n^{-a}\,\prod_{j,k}{G\({j-k\ov n}+1\)^2\ov G\({\a_j-k\ov n}+1\)\,G\({k-\a_j\ov n}+1\)}.\label{DD}\ee

Combining this with (\ref{G}) and (\ref{E}) we obtain from (\ref{result}) the main result of \cite{TW} (stated in the introduction with $t=e^{-\a}$),
\be\d(I+K_\a)\sim n^{-a}\,{\prod_{j,k}}{G\({j-k\ov n}+1\)\ov G\({\a_j-\a_k\ov n}+1\)}\,e^{-a\,\a}.\label{formreg}\ee

\begin{center}{\bf c. Remark}\end{center}

For the Toda kernels $\d(I+K_\a)$ is unchanged if we multiply  $K_a(x,y)$ by $e^{\mu(x-y)}$ for any $\mu\in \bR$. If we multiply  $K(x,y)$ by the same factor the symbol $\s(\x)$ becomes $\s(\x-i\mu)$. If this satisfies (\ref{index}), or equivalently if the index of $h(s)$ along $\Re s=\mu$ is zero, then we may apply the above results to this operator provided conditions (\ref{cond0})--(\ref{cond2}) continue to hold when the kernels $K$ and $K_\pm$ are also multiplied by $e^{\mu(x-y)}$. This is so whenever $|\mu|<\hf$. We find in the periodic case that (\ref{DD}) continues to hold when the $\a_k$ are now the zeros of $h(s)$ in the strip $\mu<\Re s<n+\mu$.

\setcounter{equation}{0}\renewcommand{\theequation}{3.\arabic{equation}}
 
\begin{center}{\bf III. The singular case}\end{center}
\begin{center}{\bf 1. Satement of the result and outline of the proof}\end{center}

We have the same setting as before, although conditions (\ref{cond0})--(\ref{cond2}) will be strengthened, but now the symbol $\s$ will be of the form
\[\s(\x)={\x^2\ov 1+\x^2}\,\t(\x),\]
where $\t$ satisfies $(\ref{index})$. 

We set
\[c^2=\lim_{\x\to0}{\s(\x)\ov \x^2}.\]
The first-order formula we shall obtain is
\[\det\,(I+K_\a)\sim 2\a\,G(\s)^{2\a}\,F(\s)\,
\]
\be\times \det(I+W^+(\s)\inv\,L_-)\,\det(I+W^-(\s)\inv\,L_+),\label{result2}\ee
where 
$W^\pm(\s)\inv$ are the inverses of $W^\pm(\s)$ acting between certain weighted $L^2$ spaces, where $G(\s)$ is the constant in the Kac-Ahiezer theorem suitably interpreted, and where
\[F(\s)=c\inv\,\exp\left\{{1\ov2\pi i}\int_{-\iy-0i}^{\iy-0i}(\log\s_+)'\,\log\s_-\,d\x\right\}\]
\[=c\inv\,
\exp\left\{-{1\ov2\pi i}\int_{-\iy+0i}^{\iy+0i}\log\s_+\,(\log\s_-)'\,d\x\right\}.\footnote{The integrals are interpreted as principal values. Although $c$ is only defined up to sign, changing the sign of $c$ changes those of $\s_\pm$ with the result that $F(\s)$ is independent of the choice.}\]
Here $\s_\pm$ are the Wiener-Hopf factors of $\s$ obtained from the factorization of $\t$: 
\[\s_\pm(\x)={\x\ov1\mp i\x}\,\t_\pm(\x),\]
normalized so that
\[\s_\pm(\x)\sim\mp i\, c \,\x\]
as $\x\to0$.

The quotient of the two sides equals $1+\Oda$ times  
\[1+p_1\,\a\inv+p_2\,\a^{-2}+\cdots,\]
where the series is an asymptotic expansion, which is also convergent for large $\a$. 

The product of determinants appearing in the formula has an integral representation for the Toda operators, and in the periodic case there are explicit representations for constants in terms of the zeros of $h(s)$. In particular $p_1$ is be given by (\ref{p1}).

Here are the conditions we assume hold in the singular case, beyond $\t$ satisfying (\ref{index}). First we assume that $e^{\dl\,|x|}k(x)\in L^1\cap\widehat{L^1}$ for some $\dl>0$.\footnote{We shall consistently use $\dl$ to denote some positive quantity, different for each occurrence.} This implies in particular that for some $\dl>0$ the operators on $L^2(\R^+)$ with kernels $e^{\dl\,(x+y)}\,k(x+y)$ and $e^{\dl\,(x+y)}\,k(-x-y)$ are trace class, and so
\be\cm\,K\,\ch_{(\a,\iy)}=O_1(e^{-\dl\a}),\ \ \ \cp\,K\,\ch_{(-\iy,-\a)}=O_1(e^{-\dl\a}).\label{cond3}\ee
(We shall consistently use $\dl$ to denote a small positive quantity, different with each appearance.) 
Second, we assume that $e^{\dl\,(|x|+|y|)}L_{\pm}(x,y)$ are trace class and that the $o_1(1)$ terms in (\ref{cond0}) and (\ref{cond1}) are $O_1(e^{-\dl\,\a})$. These assumptions are more than we actually need, but they simplify the discussion.

The proof begins as in the regular case with the approximation (\ref{Kaappr}) to $K_\a$. With our strengthened conditions the error $o_1(1)$ is sharpened to $O_1(e^{-\dl\a})$. The next step is to multiply by $\W(\s)\inv$, leading to (\ref{Kaappr2}). We shall show that in the present case
\be{\rm the\ norm\ of\ }\W(\s)\inv\ {\rm is}\ O(\a^2).\label{bound}\ee
Granting this, (\ref{Kaappr2}) becomes
\be\twotwo{\cm\,W_\a\inv\,\cm}{\cm\,W_\a\inv\,\cp}{\cp\,W_\a\inv\,\cm}{\cp\,W_\a\inv\,\cp}\,\twotwo{\cm\,T_{-\a}\,L_-\,T_{\a}\,\cm}{0}{0}
{\cp\,T_{\a}\,L_+\,T_{-\a}\,\cp}+O_1(e^{-\dl\a})\label{Kaappr3}\ee
with a different $\dl$.

The next step would be to show that the operators (\ref{first}) and (\ref{second}) converge in trace norm, the second to zero. This isn't true now.  But it will be if we think of the operators as acting on an appropriate weighted $L^2$ space. Precisely, for any $a\in \bR$ we define
\[L^2_a=\{f:e^{a|x|}\,f(x)\in L^2\}\]
with norm
\[\|f\|_a=\|e^{a|x|}\,f(x)\|,\]
the norm on the right being the usual $L^2$ norm. The space on which the operators in (\ref{Kaappr3}) act will be $L^2_{-\dl_0}$ with $\dl_0$ a sufficiently small positive quantity. Because of this change the error term $O_1(e^{-\dl\a})$ in (\ref{Kaappr3}) must be changed to $O_1(e^{(\dl_0-\dl)\a})$. If $\dl_0<\dl$ then this is still $O_1(e^{-\dl\a})$ with a different $\dl$. It is the strengthened condition on $L_\pm$ that allows us to use this weighted space: they are trace class operators from $L^2_{-\dl_0}$ to $L^2_{\dl_0}$.

To follow the argument in the regular case we also need the asymptotics of the determinant of $\W(\s)$ and the asymptotics of its inverse. We need the former for an obvious reason and the latter to determine the behavior of the operators in (\ref{first}) and (\ref{second}). The first-order asymptotics of $\d\W(\s)$ have been known for some time \cite{M}, and a first-order result for the latter is implicitly contained in \cite{W0}. But these are not enough for us even to get the first-order asymptotics here, and we want to go beyond first-order asymptotics. For application to this problem we have recently obtained sharper results \cite{W3}, which we describe in the next section. 
\pagebreak

\begin{center}{\bf 2. Asymptotic formulas for $W_\a(\s)$}\end{center}

{}In this section, where we state the results of \cite{W3}, the operators $\W(\s)$ denote $I+\ch_{(0,\a)}\,K\,\ch_{(0,\a)}$ acting on $L^2(0,\a)$. Recall that the factors $\s_\pm$ are normalized so that
\[\s_\pm(\x)\sim\mp i\, c \,\x\]
as $\x\to0$. This normalization will simplify formulas to come.

The asymptotic formula for $\d\W(\s)$ is
\be{\d\,\W(\s)\ov G(\s)^\a}\eq A\,F(\s),\label{singdet}\ee
where 
\[A= \a+i\,\({\s_+\ov\s_-}\)'(0)\]
and the sign $\eq$ indicates that the difference of the two sides is $\Oda$.
 
Stating the formula for $\W(\s)\inv$ is not as easy, and we begin with some general notation. If $\s$ is any bounded function then the Wiener-Hopf operator $W(\s)$ on $L^2(\bR^+)$ takes a function $f$ (thought of as extended by zero to $\bR^-$) with Fourier transform $\hat f$ to the inverse Fourier transform  of $\s(\x)\,\hat f(\x)$, restricted to $\bR^+$. The Hankel operator $H(\s)$ takes a function $f$ with Fourier transform $\hat f$ to the inverse Fourier transform of $\s(\x)\,\hat f(-\x)$, restricted to $\bR^+$. If $\s=\hat k$ then $W(\s)$ is the operator with kernel $k(x-y)$ while $H(\s)$ has kernel $k(x+y)$. If $\s$ extends to a bounded analytic function in the lower half-plane then $H(\s)=0$. 

The result will be expressed in terms of the inverses $\s_\pm(\x)\inv$ and Wiener-Hopf operators associated with them. These need explanation. As functions, we have
\[{1\ov\s_-(\x)}={1\ov ic\x}+u_-(\x),\ \ \ { 1\ov\s_+(\x)}={1\ov -ic\x}+u_+(\x),\]
where $u_\pm$ are bounded smooth functions. In fact we think of $\s_\pm(\x)\inv$ as distributions defined by
\be{ 1\ov\s_-(\x)}={c\inv\ov 0+i\x}+u_-(\x),\ \ \ {1\ov\s_+(\x)}={c\inv\ov 0-i\x}+u_+(\x).\label{sinv}\ee
Since $(0-i\x)\inv$ is the Fourier transform of $\cp=\ch_{\bR^+}$, we define $W((0-i\x)\inv)$ to be convolution by $\cp$ on $\bR^+$, which is defined on any locally integrable function
since
\[(\cp*f)(x)=\int_0^x f(y)\,dy.\]
Our assumption on $\s$ implies that $u_+$ is the Fourier transform of an exponentially decaying function,\footnote{Our assumtion implies that $\s_+(\x)$ extends analytically to a strip around the real line and that $\s_+(\x-i\dl)\inv$ is in the Wiener algebra (functions of the form $a+\hat k$ with $k\in L^1$) for sufficientlly small $\dl>0$. Therefore, since $(0-i(\x-i\dl))\inv=-(\dl+i\x)\inv$ is in the Wiener algebra, so is $u_+(\x-i\dl)$. This implies that $u_+$ is the Fourier transform of a function which is $O(e^{-\dl x})$.}
so $W(u_+)$ is defined on any function of at most polynomial growth. This extends the domain of $W(\s_+\inv)$ to any function of at most polynomial growth. To define $W(\s_-\inv)$ we use the fact that the Fourier transform of $(0+i\x)\inv$ is $\cm=\ch_{\bR^-}$, and convolution by this is defined only for functions in $L^1$ functions since
\[(\cm*f)(x)=\int_x^\iy f(y)\,dy.\]
Thus $W(\s_-\inv)$ acts on $L^1(\bR^+)$. 

Here is more notation. The sign $\eq$ between two operators acting on functions on $(0,\,\a)$ indicates that the difference is an integral operator with kernel having a uniform bound $\Oda$. For functions $g$ and $h$ the product $g\tn h$ denotes the operator $f\to g\,(h,f)$. For any function $\s(\x)$ we define $\tilde{\s}(\x)=\s(-\x)$. The operator $P_\a$ denotes extension by zero to $(\a,\iy)$ or restriction to $(0,\a)$, depending on the context. The operator $Q_\a$ takes a function $f(x)$ defined on $(0,\a)$ to $f(\a-x)$, and then extends it by zero to $(\a,\iy)$, or it takes a function $f(x)$ defined on $\bR^+$ to the function $f(\a-x)$ defined on $(0,\a)$. 

The result is that under the stated assumptions 

\[\W(\s)\inv\eq P_\a\,W(\s_+\inv)\,W(\s_-\inv)\,P_\a
-Q_\a\,H(\wt{u_-})\,H(u_+)\,Q_\a\]
\be-A\inv\,[Q_\a\,H(\wt{u_-})1+P_\a\,W(\s_+\inv)1]\tn[Q_\a\,H(u_+)1+P_\a\,W(\wt{\s_-}\inv)1].\label{singinv}\ee

The error term has norm $\Oda$ as an operator on $L^2$. It follows from (\ref{sinv}) that $W(\s_+\inv)$ is a bounded operator plus the operator with kernel $c\inv\,\cp(x-y)$ while $W(\s_-\inv)$ is a bounded operator plus the operator with kernel $c\inv\,\cm(x-y)$, each of which when acting on $L^2(0,\a)$ has norm $O(\a)$. The product of the operators with kernels $\cp(x-y)$ and $\cm(x-y)$, in that order, has kernel $\min(x,y)$, which has norm $O(\a^2)$. The second operator on the right is uniformly bounded, and the last with its factor $A\inv$ has norm $O(\a)$. It follows that $\|\W(\s)\inv\|=O(\a^2)$, which is (\ref{bound}).

Even now we have
\[W(\s_+\inv)\,W(\s_-\inv)\,W(\s)=W(\s_+\inv)\,W(\s_-\inv)\,W(\s_-)\,W(\s_+)=I,\]
and similarly $W(\s)\,W(\s_+\inv)\,W(\s_-\inv)=I$. These $I$ act on any spaces $L^2_{\dl_0}$, and so we are justified in writing \[W(\s_+\inv)\,W(\s_-\inv)=W(\s)\inv.\]
These operator, though, take a space $L^2_{\dl_0}$ to $L^2_{-\dl_0}$.

\begin{center}{\bf 3. Proof of (\ref{result2})}\end{center}

We begin by determining the limiting forms of the operators (\ref{first}) and (\ref{second}). As stated, we think of our operators as acting on $L^2_{-\dl_0}$ for some sufficiently small $\dl_0$, and use (the equivalent of) our assumption that $L_-$ is a trace class operator from $L^2_{-\dl_0}$ to $L^2_{\dl_0}$. 

First (\ref{first}). The summand on the left converges in trace norm to $\cm\,L_-$. Let us look at (\ref{singinv}) with $\a$ replaced by $2\a$ to see about \be\ch_{(0,\a)}(I+\ch_{(0,2\a)}\,K\,\ch_{(0,2\a)})\inv\,\ch_{(0,\a)}.\label{seeabout}\ee
The last term in (\ref{singinv}) has norm $O(\a\inv)$ as an operator from $L^2_{\dl_0}$ to $L^2_{-\dl_0}$, while the first term convrges strongly to $W(\s)\inv$. The kernel of the second term, the one with the $Q_\a$, after multiplying by $e^{-\dl_0(x+y)}$, becomes $\Oda$.\footnote{Because $e^{-\dl_0(x+y)}\,e^{-\dl(2\a-x-y)}\le e^{-2\dl\a}$ if $\dl<\dl_0$. Recall that $u_+$ and $\tl u_-$ are Fourier transforms of exponentially decaying functions.} Therefore as an operator from $L^2_{\dl_0}$ to $L^2_{-\dl_0}$ has norm $\Oda$. So the product with $L_-\,\ch_{(-\iy,\a)}$ converges in trace norm to $\cp\,W(\s)\inv\,\cp\,L_-$. Hence if we define $W^+(\s)$ to be the direct sum of $W(\s)\inv$ on $\bR^+$ and $I$ in $\bR^-$ (it is the inverse of $I+\cp\,K\,\cp$ in an appropriate sense) we see that (\ref{first}) converges in trace norm to $W^+(\s)\inv\,L_-$. 
In fact (\ref{singinv}) gives the sharper formula for (\ref{seeabout}),
\be W^+(\s)\inv-A\inv\,W(\s_+\inv)1\tn W(\s_-\inv)1+\Oda,\label{sharper}\ee
where the error refers to the norm as an operator from $L_{\dl_0}^2$ to $L^2_{-\dl_0}$, with a corresponding sharper formula for (\ref{first}).

For (\ref{second}) we change $\a$ to $\a/2$ to conform to the notation of the preceding paragraphs and so write it as
\be\ch_{(-\a/2,0)}\,T_{-\a}\,\W(\s)\inv\,
\ch_{(0,\a/2)}\,L_-\,\ch_{(-\iy,\a/2)}.\label{secondsecond}\ee
The factor $L_-\,\ch_{(-\iy,\a/2)}$ converges in trace norm to $L_-$. The remaining factor  is equal to 
\[U\,\ch_{(0,\a/2)}\,Q_\a\,\W(\s)\inv\,\ch_{(0,\a/2)},\]
where $U$ is the unitary operator defined by $Ug(x)=g(-x)$. If we multiply (\ref{singinv}) on the left by $Q_\a$ we obtain
\[Q_\a\,W(\s_+\inv)\,W(\s_-\inv)\,P_\a
-P_\a\,H(\wt{u_-})\,H(u_+)\,Q_\a\]
\[-A\inv\,[P_\a\,H(\wt{u_-})1+Q_\a\,W(\s_+\inv)1]\tn[Q_\a\,H(u_+)1+P_\a\,W(\wt{\s_-}\inv)1].\]
Removing the operators which are $\Oda$ gives
\pagebreak
\[Q_\a\,W(\s_+\inv)\,W(\s_-\inv)\,P_\a-A\inv\,Q_\a\,W(\s_+\inv)1\tn P_\a\,W(\wt{\s_-}\inv)1\]
\[-A\inv\,P_\a\,H(\wt{u_-})1\tn P_\a\,W(\wt{\s_-}\inv)1.\]
We have
\[ W(\s_+\inv)1=c\inv x+O(1),\]
and so 
\[Q_\a\,W(\s_+\inv)1=c\inv \a+O(x),\]
so the second term above equals 
\be-c\inv\,1\tn W(\wt{\s_-}\inv)1\label{1tnW}\ee
plus an operator acting between our weighted spaces with norm $O(\a\inv)$. The last term has norm $O(\a\inv)$. As for the first,
the operator $W(\s_+\inv)$ is $c\inv\,\cp(x-y)+W(u_-)$, to abuse notation, while $W(\s_-\inv)$ is
 $c\inv\,\cm(x-y)+W(u_-)$, so if $k_\pm$ denote the inverse Fourier transforms of $u_\pm$ the product is the operator with kernel
\[c^{-2}\,\min(x,y)+c\inv\int_0^x k_-(z-y)\,dz+c\inv\int_0^y k_+(x-z)\,dz+\int_{y<z<x}k_+(x-z)\,k_-(z-y)\,dz.\]
To apply $Q_\a$ we replace $x$ by $\a-x$, and then we take $x,\,y\in(0,\a/2).$ The first term  becomes $c\inv\,y$. The second term becomes
\[c\inv\int_0^{\a-x} k_-(z-y)\,dz=c\inv\int_0^\iy k_-(z-y)\,dz,\]
since $\a-x>y$ when $x,\,y\in(0,\a/2).$ Similarly the last term equals zero. The remaining term becomes
\[c\inv\int_0^y k_+(\a-x-z)\,dz=c\inv\int_{\a-x-y}^{\a-x} k_+(z)\,dz.\]
If we multiply this by $e^{-\dl(x+y)}$ this becomes $\Oda$ when $x,\,y\in(0,\a/2).$ Thus the entire expression is $\Oda$ plus the operator with kernel
\[c\inv\,\left[c\inv\,y+\int_0^\iy k_-(z-y)\,dz\right].\]
This is precisely the kernel of (\ref{1tnW}) without its minus sign. 
This shows that (\ref{secondsecond}) has trace norm $O(\a\inv)$, and therefore so does (\ref{second}).

With the analogous results for the two other matrix entries, we have shown that the operator (\ref{Kaappr3}) converges in the trace norm of operators on $L^2_{-\dl_0}$ to
\[\twotwo{I+W^+(\s)\inv\,L_-}{0}{0}{I+W^-(\s)\inv\,L_+},\footnote{The derivation shows that the correction is of the form $A\inv T+O_1(e^{-\dl\a})$ for some operator $T$. The upper-left corner of $T$ is the second summand in (\ref{sharper}). We shall use this, and its analogue for the lower-right corner, when we obtain more explicit formulas. For now we observe that this and (\ref{singdet}) imply that the ratio of the two sides in the asymptotics of $\d(I+K_\a)$ has the form stated in the introduction}\]
and so the determinant of $I$ plus the operator converges to \[\det(I+W^+(\s)\inv\,L_-)\,\det(I+W^-(\s)\inv\,L_+).\]

Together with (\ref{singdet}), with $\a$ replaced by $2\a$ since the underlying interval is $(-\a,\,\a)$ rather than $(0,\,\a)$, this establishes (\ref{result2}).

\begin{center}{\bf 4. The Toda operators}\end{center}

\begin{center}{\bf a. Evaluation of the determinants in (\ref{result2})}\end{center}

Just as in the regular case it is possible to get more concrete formulas for the product of determinants in (\ref{result2}) for the Toda operators. The main point is that it is the limit of the corresponding products in the regular case.  

Consider $\s_\l$ with $\l$ close to 1, but not equal to 1, with ${\rm arg}\,(1-\l)={\rm arg}\,c^2$.
The double zero of $\s$ at 0 splits into zeros of $\s_\l$ at
\[\pm i\sqrt{1-\l\ov\l\,c^2}.\]
We denote them by $i\ep_\pm$ with $\Re \ep_+>0,\ \Re\ep_-<0$. Corresponding to the representation 
\[\s(\x)={\x^2\ov\x^2+1}\t(\x)\]
with factors
\[\s_\pm(\x)={\x\ov \x\pm i}\t_\pm(\x)\]
there is a representation
\[\s_\l(\x)={(\x-i\ep_-)(\x-i\ep_+)\ov\x^2+1}\t_\l(\x)\]
with factors
\[{\s_\l}_\pm(\x)={\x-i\ep_\mp\ov \x\pm i}{\t_\l}_\pm(\x).\]
 
Because of the formulas
\[W(\s_\l)\inv=W({\s_\l}_+\inv)\,W({\s_\l}_-\inv),\ \ \ 
W(\s)\inv=W(\s_+\inv)\,W(\s_-\inv),\]
it is straightforward to show that as $\l\to1$ $W(\s_\l)\inv$ converges to $W(\s)\inv$, as operators from $L^2_{\dl_0}$ to $L^2_{-\dl_0}$. This shows that the determinants in (\ref{result2}) for $\s_\l$ converge to those for $\s$. In particular we see that the formula (\ref{detprodrep}) for the Toda operators continues to hold in the singular case.

In the periodic subcase fomula (\ref{DD}) continues to hold when suitably interpreted. With $\l$ near 1 the zeros $i\ep_\pm$ for $\s_\l$ correspond for $h_\l$ to two zeros in the strip 
\linebreak
$\hf<|\Re s|<n+\hf$, one near $\hf$ and one near $n+\hf$. Thus in the limit one $\a_j$ becomes $\hf$ and one becomes $n+\hf$, and that is how (\ref{DD}) is interpreted in the singular case.

Next we compute $F(\s)$. With $\s_\l$ as before we have
\[\log E(\s_\l)={1\ov2\pi i}
\int_{-\iy}^{\iy}{{\s_\l}_+'(\x)\ov {\s_\l}_+(\x)}\,\log{\s_\l}_-(\x)\,\,d\x\]
\[={1\ov2\pi i}
\int_{-\iy-\dl i}^{\iy-\dl i}{{\s_\l}_+'(\x)\ov {\s_\l}_+(\x)}\,\log{\s_\l}_-(\x)\,\,d\x-\log {\s_\l}_-(\ep_-),\]
where we used the fact that ${\s_\l}_+(\x)$ has a zero at $\x=i\ep_-$. Here $\dl$ is small, positive, and fixed. This may be written
\[\log E(\s_\l)+\log{\ep_--\ep_+\ov\ep_--1}={1\ov2\pi i}
\int_{-\iy-\dl i}^{\iy-\dl i}{{\s_\l}_+'(\x)\ov {\s_\l}_+(\x)}\,\log{\s_\l}_-(\x)\,\,d\x-\log {\t_\l}_-(\ep_-).\]

Now our normalizations of $\s_\pm$ give $\t_\pm(0)=c$, and ${\t_\l}_\pm\to \t_\pm$ uniformly near $\x=0$ as $\l\to1$. Hence
as $\la\to1$ the right side above tends to $\log F(\s)$  and so we deduce that
\[F(\s)=\lim_{\la\to1} \,(\ep_-+\ep_-)\,E(\s_\l).\]
The zeros of $\s_\l$ at $i\ep_\pm$ correspond, under the variable change $s=\hf-i\x$, to zeros of $h_\l$ at $\hf+\ep_+$ and $\hf+\ep_-$. The latter gives the zero at $\hf+\ep_-+n$ in the strip $\hf<|\Re s|<n+\hf$. If we look at the terms in the expression for $E(\s_\l)$ we see that as $\l\to1$ all terms tend to nonzero limits except for the first factor in the denominator when $\a_j=\hf+\ep_+$ and $\a_k=\hf+\ep_-+n$. For these $j$ and $k$
\[\lim_{\la\to1}{\ep_+-\ep_-\ov G\({\a_j-\a_k\ov n}+1\)}=
\lim_{\la\to1}{\ep_+-\ep_-\ov G\({\ep_+-\ep_-\ov n}\)}=n.\]
Thus 
\[F(\s)=n\,{\prod_{j,k}}'{G\({\a_j-k\ov n}+1\)\ov G\({\a_j-\a_k\ov n}+1\)}
{G\({j-\a_k\ov n}+1\)\ov G\({j-k\ov n}+1\)},\]
where the prime indicates that the factor $G\({\a_j-\a_k\ov n}+1\)$ coming from the pair of zeros $\a_j=\hf,\ \a_k=\hf+n$ is removed from the denominator.

{}From what we have done we see that the constant factor
\[2\,F(\s)\,\d(I+(W^+)\inv\,L_-)\,\d(I+(W^-)\inv\,L_+ )\]
in the asymptotic formula is equal to
\[2\,n^{1-a}\,{\prod_{j,k}}'{G\({j-k\ov n}+1\)\ov G\({\a_j-\a_k\ov n}+1\)},\]
the prime having the same meaning as before, and so
\be\d(I+K_\a)\sim 2\,n^{1-a}\,{\prod_{j,k}}'{G\({j-k\ov n}+1\)\ov G\({\a_j-\a_k\ov n}+1\)}\,\a\,e^{-a\,\a},\label{formsing}\ee
which is the formula stated in the introduction after setting $t=e^{-\a}$.

\begin{center}{\bf b. The next approximation}\end{center}

The $\sim$ sign in (\ref{result2}) may be replaced by a series
\[1+p_1\,\a\inv+p_2\,\a^{-2}+\cdots,\]
which is both asymptotic and convergent for large $\a$. This is clear from the expressions with error $\Oda$  we have obtained for $\d\W(\s)$ and for (\ref{first}) and (\ref{second}) and its analogues. Here we compute the coefficient $p_1$. We begin by using the sharp asymptotics of $\W(\s)\inv$. 

What appears in (\ref{first}) and (\ref{second}) is not the operator $(I+\ch_{(0,\a)}\,K\,\ch_{(0,\a)})\inv$ but $(I+\ch_{(0,2\a)}\,K\,\ch_{(0,2\a)})\inv$. So from (\ref{sharper}) we see that $I$ plus (\ref{prod}) equals 
\[I+\twotwo{W^+(\s)\inv\,L_-}{0}{0}{W^-(\s)\inv\,L_+}\]
\be B\inv\,
\twotwo{-W(\s_+\inv)\,(\cp\tn\cp)\,W(\s_-\inv)\,L_-}{\times}{\times}{\times}+O_1(e^{-\dl\a}),\label{2nd}\ee
where  
\[B=2\a+i\,\({\s_+\ov\s_-}\)'(0).\]
The crosses in the last matrix denote certain operators; the lower-right entry is analogous to the upper-left entry. The operator $W(\s_+\inv)\,(\cp\tn\cp)\,W(\s_-\inv)$ may be thought of as having the factor $\cp$ on both sides. The reason is that the second approximation to the direct summand of 
$(I+\ch_{(0,\a)}\,K\,\ch_{(0,\a)})\inv$ acting on $L^2(\R^-)$ is zero. To get the second approximation to the determinant we factor out the sum of the first two operators. The determinant of the result is to first order the trace of the second summand which, because we factored out a diagonal matrix, is just the sum of the traces of $(I+W^\pm(\s)\inv\,L_\mp)\inv$ times the diagonal entries of the last matrix. Thus we want to compute the trace of 
\[-(I+W^+(\s)\inv\,L_-)\inv\,W(\s_+\inv)\,(\cp\tn\cp)\,W(\s_-\inv)\,L_-\]
and its analogue from the lower-right corner. 

This is completely general. To evaluate this for the Toda operators we shall, as we did before, perturb $\s$ so that $W(\s)$ becomes invertible in the usual sense, evaluate the trace in this case, and then take the limit. So for now let us assume $W(\s)$ is invertible.

Keeping in mind the factors $\cp$ we see that the displayed operator may be written
\[-(W^+(\s)+L_-)\inv\,W(\s)\,W(\s_+\inv)\,
(\cp\tn\cp)\,W(\s_-\inv)\,L_-.\]
Since $W(\s)=W(\s_-)\,W(\s_+)$ and $W(\s_-)1=\s_-(0)1$ by an earlier argument, we see that the last operator equals 
\[-\s_-(0)\,(W^+(\s)+L_-)\inv\,(\cp\tn\cp)\,W(\s_-\inv)\,L_-,\]
which has the same trace as
\[-\s_-(0)\,L_-\,(W^+(\s)+L_-)\inv\,(\cp\tn\cp)\,W(\s_-\inv)\]
\[=\s_-(0)\,\Big(W^+(\s)\,(W^+(\s)+L_-)\inv-I\Big)\,(\cp\tn\cp)\,W(\s_-\inv).\]
This has the same trace as what results when we left-multiply by $W^+(\s)\inv$ and right-multiply by $W^+(\s)$. Using
\[(\cp\tn\cp)W(\s_-\inv)\,W(\s)=
(\cp\tn\cp)\,W(\s_+)=\cp\tn(W(\widetilde{\s_+})\cp)=\s_+(0)\,\cp\tn\cp,\]
we see that what results from these multiplications is
\[\s(0)\Big((W^+(\s)+L_-)\inv-W^+(\s)\inv\Big)\,\cp\tn\cp
=\s(0)\Big((I+K_-)\inv-W^+(\s)\inv\Big)\,\cp\tn\cp.\]

Recall the notation of  \S II.3.a, where  $K_-=A\cp B$ while $W^+(\s)=I+\cp\,AB\,\cp$. (As in that section we may assume that the support of $\r$ lies in the left half-plane and modify $K_\pm$.) We can left-multiply the displayed operator by $\cp$ without changing the trace and then the operator on the left of the tensor product becomes, aside from the factor $\s(0)$, 

\[I-W(a)\,W(1+ab)\inv\,W(b)-W(1+ab)\inv\]
\be=I-W(\s)\inv-W(a)\,W(\s)\inv\,W(b).\label{WW}\ee
A manipulation shows that this equals
\[ H\({1\ov\s_+}\)\,H\({1\ov\wt{\s_-}}\)+H\({a\ov\s_+}\)\,H\({\tl b\ov\wt{\s_-}}\).\]
The trace of this is evaluated using the formula
\[{\rm tr}\; H(\ph)\,H(\tl{\ps})1\tn1={1\ov2\pi}\int_{-\iy}^\iy
\({\ph(\x)\ov0+i\x}\)_+\,\({\ps(\x)\ov0-i\x}\)_-\,d\x.\]
This is equal to
\be {1\ov2\pi}\int_{-\iy}^\iy
\({\ph(\x)\ov0+i\x}\)_+\,{\ps(\x)\ov0-i\x}\,d\x.\label{traceint}\ee

For the case in hand our perturbed $\s$ will have zeroes $i\ep_+$ in the upper half-plane and $i\ep_-$ in the lower. Consider first the integral correponding to $\ph=\s_+\inv,\ \ps=\s_-\inv$. We want to move the contour of integration into the upper half-plane, where the integral remains bounded as we unperturb $\s$. We pass through the pole of the second factor at $i\ep_+$ so the above equals $O(1)$ plus
\[{i\ov\ep_+}\,{\rm Res}(\s_-\inv,i\ep_+)\,\({\s_+(\x)\inv\ov0+i\x}\)_+(i\ep_+).\]
The last factor is equal to
\[{1\ov2\pi i}\int_{-\iy}^{\iy}{\s_+(\eta)\inv\ov 0+i\eta}\,
{d\eta\ov\eta-i\ep_+}.\]
We move this contour into the lower half-plane where the resulting integral will be $O(1)$. We pass through the pole at $\eta=i\ep_-$ whose contribution is
\[{1\ov\ep_-}\,{\rm Res}(\s_+\inv,i\ep_-)\,{1\ov i(\ep_--\ep_+)}.\]
Thus in the case $\ph=\s_+\inv,\ \ps=\s_-\inv$ (\ref{traceint}) equals $O(1)$ plus
\[{1\ov \ep_-\,\ep_+(\ep_--\ep_+)}\,{\rm Res}(\s_-\inv,i\ep_+)\,
{\rm Res}(\s_+\inv,i\ep_-).\]

When $\ph=\l\,a\,\s_+\inv,\ \ps=b\,\s_-\inv$ (recall that we are really dealing with $\s_\l$) the only difference is in the residues, and so the sum of the two equals
\[{1\ov \ep_-\,\ep_+(\ep_--\ep_+)}\,{\rm Res}(\s_-\inv,i\ep_+)\,
{\rm Res}(\s_+\inv,i\ep_-)\,(1+\l\,a(i\ep_-)\,b(i\ep_+)).\]
Since $1+\l\,a(i\ep_-)\,b(i\ep_-)=\s_\l(i\ep_-)=0$ we can replace the last factor by
\[\l\,a(i\ep_-)\,(b(i\ep_+)-b(i\ep_-))\sim i\,a(0)\,b'(0)\,(\ep_+-\ep_-)\]
as $\l\to1$. Notice that in the limit $0=\s(0)=1+a(0)\,b(0)$ so $a(0)=-1/b(0)$. 

At this point we have shown that as $\l\to1$ the trace of the operator (\ref{WW}) times $1\tn1$ is asymptotically equal to
\be{i\ov \ep_-\,\ep_+}\,{\rm Res}(\s_-\inv,i\ep_+)\,
{\rm Res}(\s_+\inv,i\ep_-)\,{b'(0)\ov b(0)}.\label{point}\ee

Recall that we are computing the trace of
\[-(I+W^+(\s)\inv\,L_-)\inv\,W(\s_+\inv)\,(1\tn1)\,W(\s_-\inv)\,L_-,\]
which is equal to $\s(0)$ times the trace of (\ref{WW}) times $1\tn1$. So we have to find the limit as $\l\to1$ of $\s(0)$ times (\ref{point}). There is a factorization
\[\s(\x)=(\x-i\ep_-)\,(\x-i\ep_+)\,v(\x),\]
where $v$ is nonzero with factors $v_\pm$.\footnote{In the notation of \S III.4.a $v(\x)=\t_\l(\x)/(\x^2+1)$, and the corresponding factors are $v_\pm(\x)=(\t_\l)_\pm(\x)/(\x\pm i)$.}  We have
\[\s(0)=-\ep_-\,\ep_+\,v(0),\ \ \ {\rm Res}(\s_-\inv,i\ep_+)=v_-(i\ep_+)\inv,\ \ \ 
{\rm Res}(\s_+\inv,i\ep_-)=v_+(i\ep_-)\inv.\]
In the limit the product of the residues cancels the factor $v(0)$ in the expression for $\s(0)$. This shows that for our unperturbed~$\s$ the trace in question is equal to
\be-i{b'(0)\ov b(0)}={\G'\ov \G}\!\(\hf\)=-\ga-2\,\log 2,\label{ga}\ee
since $b(\x)=\G(\hf+i\x)$.

We have to divide by 2 (recall the $2\a$ in the definition of $B$) and then add the contribution of the lower-right corner, which will be the same. The upshot is that the cofficient of $\a\inv$ in the expansion of the determinant of (\ref{2nd}) is (\ref{ga}) independently of $\s$. This is its contribution to the coefficient $p_1$.

Finally we compute the contribution to $p_1$ of $\d\W(\s)$. Of course this is
\[-{i\ov2}\,\({\s_+\ov\s_-}\)'(0),\]
and in general nothing more can be said. But we can compute it explicitly in the periodic case. 

Since $(\s_+/\s_-)(0)=-1$, it is the same as
\[{i\ov2}\,{d\ov d\x}\log{\s_+(\x)\ov\s_-(\x)}\Big|_{\x=0}.\]
Again we shall use the variable $s=\hf-i\x$ instead of $\x$, with factors $h=h_-\,h_+$ corresponding to $\s=\s_-\,\s_+$, and the above equals
\be-{1\ov2}\,{d\ov ds}\log{h_+(s)\ov h_-(s)}\Big|_{s={1\ov2}}.\label{hh}\ee

{}From \cite[p.~217]{TW} we have the explicit factorization
\[ h_-(s)={(-1)^n\,2^{n-1}\,\G(1-s)\,n^{s}\ov\prod\G\({\a_k-s\ov n}\)},\ \ \ h_+(s)={\G(s)\,n^{-s}\ov\prod\G\({s-\a_k\ov n}+1\)}.\]
Here the $\a_k$ are the $n$ zeros of $h(s)$ in the strip $\hf\le\Re s\le n+\hf$, as in \S II.3.b. Since we are in the singular case one of the zeros, say $\a_1$, is $\hf$ while another, say $\a_n$, is $n+\hf$; the other $n-2$ zeros are in the interior of the strip. Thus $h_+(s)/h_-(s)$ equals a constant times
\[n^{-2s}\,{\G(s)\ov\G(1-s)}\,{\G\({\hf-s\ov n}\)\,\G\({\hf-s\ov n}+1\)\ov\G\({s-\hf\ov n}+1\)\,\G\({s-\hf\ov n}\)}
\,\prod_{k\ne1,\,n}{\G\({\a_k-s\ov n}\)\ov\G\({s-\a_k\ov n}+1\)}.\]
{}From this find that (\ref{hh}) is equal to
\[-{\G'\ov\G}\!\(\hf\)+\log n+{2\ov n}\,{\G'\ov\G}\!\(1\)
+{1\ov 2n}\sum_{k\ne1,n}\left[{\G'\ov\G}\!\({\a_k-\hf\ov n}\)+{\G'\ov\G}\!\({\hf-\a_k\ov n}+1\)\right]\]
\[=\(1-{2\ov n}\)\,\ga+2\,\log 2+\log n+{1\ov 2n}\sum_{k\ne1,n}\left[{\G'\ov\G}\!\({\a_k-\hf\ov n}\)+{\G'\ov\G}\!\({\hf-\a_k\ov n}+1\)\right].\]

This is the contribution of $\d\W(\s)$ to $p_1$. To this we  add (\ref{ga}), obtaining (\ref{p1}).

\begin{center}{\bf c. The cases $n=2$ and $n=3$}\end{center}

When $n=2$ the measure $\r$ is supported in the single point $-1$ and the function $q=q_0$ satisfies the cylindrical sinh-Gordon equation
\[q''(t)+t\inv\,q(t)=8\,\sinh 2q.\] 
When $\r(\{-1\})=-1/\pi$  the symbol is
$1-{\rm sech}\,\pi\x$
and this is in the singular case. In the notation of the beginning of the introduction this is the symbol for the operator $K_0$ and the symbol for $K_{-1}$ is
$1+{\rm sech}\,\pi\x$,
which is in the regular case. For $K_0$ the zeros $\a_k$ are seen to be $1/2$ and $5/2$ while for $K_{-1}$ there is a double zero at $3/2$. Using formulas (\ref{formsing}), (\ref{formreg}) and (\ref{p1}) and recalling that $\a=\log t\inv$ we obtain 
\[e^{q(t)}=2\,t\,\Big(\log t\inv+\log2-\ga+O(1/\log t)\Big),\]
in agreement with \cite{MTW}. 
 
When $n=3$ and $\r(\{e^{2\pi i/3}\})+\r(\{e^{-2\pi i/3}\})=0$ the function $q=q_0$ satisfies the Bullough-Dodd equation
\[q''(t)+t\inv\,q(t)=4\,e^{2q}-4\,e^{-q},\]
and
\[h(s)=1+\l{\sin\pi((s+2)/3)\ov\sin\pi s},\]
with $\l$ a constant times $\r(\{e^{2\pi i/3}\})$. (See \cite[\S 7]{TW}.) There is no pole at $s=1$ which means that this point should count as a zero. The zeros are given by
\[\a_0(\l)={1\ov4}-{3\ov2\pi}{\rm arcsine}\({1+\l\ov2}\),\ \ \ \a_1(\l)=1,\ \ \ \a_2(\l)=2-\a_0(\l).\] 
As $\l$ increases from 0 to 1 $\a_0(\l)$ decreases to $-1/2$, $\a_1(\l)=1$, and $\a_2(\l)$ increases to $5/2$, and so when $\l=1$ $K_0$ is in the regular case with $\a_1=1,\ \a_2=5/2,\ \a_3=5/2$. For $K_{-1}$ the zeros get translated one unit to the right, so $\a_0(\l)$ increases to $1/2$, $\a_1(\l)$ decreases to $1/2$, and $\a_2(\la)=2$. Thus when $\l=1$ $K_{-1}$ is in the singular case with $\a_1=1/2,\ \a_2=2,\ \a_3=7/2$. Using formulas (\ref{formsing}), (\ref{formreg}) and (\ref{p1}) now gives 
\[e^{-q(t)}=2\,t\,\Big(\log t\inv+\log3-{2\ov3}\log 2-\ga+O(1/\log t)\Big),\]
in agreement with \cite{TW} (and with \cite{K} up to a factor 2).

There is another singular situation when $\l=-3$. As $\l$ decreases from 0 to $-3$ $\a_0(\l)$ increases to 1, $\a_1(\l)=1$, and $\a_2(\l)$ decreases to 1. Thus $K_0$ is not in one of the singular cases  covered by this work because the zeros $\a_0(\l)$ and $\a_1(\l)$ coalesce at a boundary point of the region $0<\Re s<1$. Indeed, the asymptotics of $\det(I+K_0)$ are different here \cite{K,TW}: the power of $t$ is as expected but the factor $\log t\inv$ appears to the second power rather than the first.

\begin{center}{\bf Acknowledgment}\end{center}

This work was supported by National Science Foundation grant DMS-0552388.


\begin{thebibliography}{999}

\bibitem{A}N. I. Achieser, {\it The continuous analogue of some theorems on Toeplitz matrices} (Russian), Ukrain. Mat. Zh. {\bf 16}:4 (1964) 445--462.

\bibitem{GF} I. C. Gohberg and I. F. Feldman, Convolution Equations and Projection Methods for their Solution, Transl. Math. Monog. {\bf 41}, Amer. Math. Soc., 1974.

\bibitem{K} M. Kac, {\it Toeplitz matries, translation kernels and a related problem in probability theory}, Duke Math. J. {\bf 21} (1954) 501--510.

\bibitem{Ki} A. V. Kitaev, {\it Method of isometric deformation for ``degenerate''
third Painlev\'e equation}. J. Soviet Math. {\bf 46} (1989) 2077--2082. 

\bibitem{M} L. V. Mikaelyan, {\it Asymptotics of determinants of truncated Wiener-Hopf operators in a singular case}, (Russian)  Akad. Nauk Armyan. SSR Dokl.  {\bf 82}  (1986) 151--155.

\bibitem{MTW} B. M. McCoy, C. A. Tracy, and T. T. Wu, {\it Painlev\'e functions of the
third kind}. J. Math. Phys. {\bf 18} (1977) 1058--1092. 

\bibitem{TW} C. A. Tracy and H. Widom, {\it Asymptotics of a class of solutions to the cylindrical Toda equations}, Commun. Math. Phys. {\bf 190} (1998) 697--721.

\bibitem{W0} H. Widom, {\it Extreme eigenvalues of N-dimensional convolution operators}, Trans. Amer. Math. Soc. 106 (1963) 391--414.

\bibitem{W1} H. Widom, {\it Some classes of solutions to the Toda lattice hierarchy}, Comm. Math. Phys. {\bf 184} (1997) 653--667.

\bibitem{W2} H. Widom, {\it Asymptotics of a Class of Operator Determinants}, to appear in Oper. Th.: Adv. Appl., arXiv: math.FA/0504257.

\bibitem{W3} H. Widom, {\it On the inverse and determinant of a truncated Wiener-Hopf operator}, arXiv: math.FA/0605076.


\end{thebibliography}
\end{document}